\documentclass{moriond}

\usepackage{amssymb}
\usepackage{amsmath}

\def\be{\begin{equation}}
\def\ee{\end{equation}}
\def\bea{\begin{eqnarray}}
\def\eea{\end{eqnarray}}

\newcommand{\alphas}{\alpha_\mathrm{s}}
\newcommand{\rw}{\mathswitchr w}
\newcommand{\sw}{\mathswitch {s_\rw}}

\let\gsim\gtrsim
\let\lsim\lesssim

\newcommand{\QCD}{{\mathrm{QCD}}}
\newcommand{\EW}{{\mathrm{EW}}}
\newcommand{\LO}{{\mathrm{LO}}}
\newcommand{\NLO}{{\mathrm{NLO}}}

\def\reffi#1{\mbox{Figure~\ref{#1}}}

\def\refta#1{\mbox{Table~\ref{#1}}}

\def\citere#1{\mbox{Ref.~\cite{#1}}}
\def\citeres#1{\mbox{Refs.~\cite{#1}}}

\def\mathswitchr#1{\relax\ifmmode{\mathrm{#1}}\else$\mathrm{#1}$\fi}

\newcommand{\PW}{\mathswitchr W}

\newcommand{\PZ}{\mathswitchr Z}

\newcommand{\PH}{\mathswitchr H}
\newcommand{\Pe}{\mathswitchr e}

\newcommand{\Pj}{\mathswitchr j}

\newcommand{\Pp}{\mathswitchr p}

\newcommand{\Pep}{\mathswitchr {e^+}}

\def\mathswitch#1{\relax\ifmmode#1\else$#1$\fi}

\newcommand{\MW}{\mathswitch {M_\PW}}

\newcommand{\GeV}{\unskip\,\mathrm{GeV}}

\newcommand{\rT}{{\mathrm{T}}}

\newcommand{\Photo}{}

\begin{document}
\mbox{} \hfill FR-PHENO-2026-007
\vspace*{4cm}
\title{Precision calculations for electroweak multi-boson processes
\footnote{Contribution to the Proceedings of the 60th Rencontres de Moriond
``Electroweak Interactions \& Unified Theories'', 2026.}}

\author{S. Dittmaier}

\address{University of Freiburg, Institute of Physics, Hermann-Herder-Str.3,\\
79104 Freiburg, Germany}

\maketitle\abstracts{
We review the salient features of next-to-leading-order QCD and electroweak
corrections to the scattering of two and the production of three weak gauge bosons
at the Large Hadron Collider. Results for the tower
of ${\cal O}(\alphas^m\alpha^n)$ corrections are shown for the exemplary processes
of like-sign WW~scattering and triple-W production, emphasizing the large 
impact of purely electroweak corrections which generically grow to 
$\sim-16\%$ and $\sim-7\%$ for these process types, respectively, even for
integrated cross sections.
Moreover, we discuss the possibility to reproduce the results of full off-shell 
calculations by the {\it vector-boson scattering approximation,}
{\it leading-pole approximations,} and the {\it effective vector-boson approximation.}
}

\section{Introduction}

Already since the early days of spontaneously broken gauge theories,
the process class of electroweak (EW) vector-boson scattering (VBS) 
raised particular interest owing to its simultaneous sensitivity to
gauge-boson self-interactions and to the mechanism of EW symmetry breaking (EWSB).
A similar sensitivity is provided by triple-gauge-boson production,
called VVV production in the following. 
These process classes are analyzed by ATLAS and CMS in many
different channels, up to now confirming the Standard Model (SM) predictions~\cite{Folgueras}.
Future prospects for LHC Run~3 and the HL-LHC promise experimental
sensitivities to the few-percent level in integrated cross sections
as well as the isolation of contributions of longitudinal W/Z~bosons,
which most directly probe EWSB~\cite{Han}.

\begin{figure}
\centerline{
\includegraphics[bb=130 567 485 665,width=0.7\textwidth]{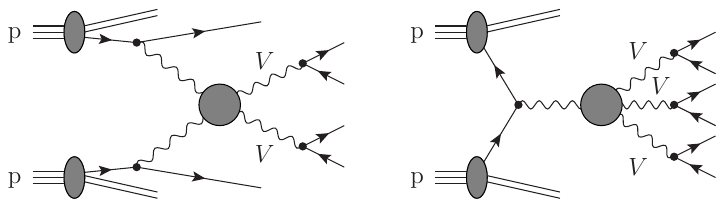}
}
\caption{Typical diagram structures for 
EW VBS (left) and VVV production (right) in $\Pp\Pp$
collisions.}
\label{fig:LO-diagrams}
\end{figure}

Figure~\ref{fig:LO-diagrams} illustrates the embedding of the most interesting
$VV\to VV$ and $V\to VVV$ subprocesses in the
VBS and VVV production mechanisms in pp~collisions.
The underlying amplitudes for given final-state fermions, however,
feature already $\sim10^2$ Feynman diagrams at leading order (LO) and 
$\sim 10^3{-}10^4$ diagrams at next-to-leading order (NLO),
where the most complicated graphs involve 8-point one-loop integrals.
For the most important VBS and VVV processes, results for the full tower
of ${\cal O}(\alphas^m\alpha^n)$ corrections exist in the 
literature (as, e.g., reviewed in 
\citeres{Covarelli:2021gyz,Huss:2025nlt}),
but the feasibility of such multi-leg NLO calculations came only after several
decades of continuous conceptual and technical progress.
Apart from the sheer algebraic complexity of the amplitudes, the
major issues concerned the gauge-invariant treatment of the W/Z~resonances,
the separation and cancellation of infrared divergences, and 
the numerically stable evaluation of the multi-leg one-loop integrals
(see, e.g., \citere{Denner:2019vbn} for a review of conceptual and technical aspects).
The results shown and discussed below have been obtained with the
dedicated Monte Carlo integrater 
{\sc Bonsay}~\cite{Denner:2019tmn,Dittmaier:2019twg,Denner:2022pwc,Dittmaier:2023nac}, 
which employs tree-level and one-loop
amplitudes from the matrix-element generators 
{\sc Openloops2}~\cite{Buccioni:2019sur} or
{\sc Recola}~\cite{Actis:2012qn}
and evaluates one-loop integrals with {\sc Collier}~\cite{Denner:2016kdg},
or with the Monte Carlo program
{\sc MoCaNLO}~\cite{Denner:2026phn}, which is based on {\sc Recola} and {\sc Collier}.
Similar NLO results should, for instance, be producible with 
{\sc MadGraph5\_aMC@NLO}~\cite{Frederix:2018nkq}
or {\sc Sherpa}~\cite{Gleisberg:2008ta} as well, however, the level of complexity
of the underlying computations certainly calls for careful cross-checks of results
obtained with different tools.

\section{Electroweak vector-boson scattering}

EW VBS, $VV\to VV$, is mostly analyzed via final states with
leptonically decaying EW gauge bosons, so that the experimental signature comprises
four leptons and two jets ($4\ell2\Pj$).
The contributions to the LO cross section for $4\ell2\Pj$ production scale like
${\cal O}(\alpha^6)$, ${\cal O}(\alphas^2\alpha^4)$, and ${\cal O}(\alphas\alpha^5)$
for the purely EW graphs, squared graphs with initial-state gluons or gluon exchange, and
the interferences between EW and gluon-exchange graphs, respectively.
Apart from the like-sign VBS channel with $\PW^\pm\PW^\pm$ intermediate states,
the respective final states with four leptons and two quarks receive large LO
contributions of ${\cal O}(\alphas^2\alpha^4)$ from gluon fusion.
Note also that all VBS channels receive LO contributions from VVV production.
In order to enhance the purely EW VBS contribution of ${\cal O}(\alpha^6)$ to the
overall cross section, dedicated VBS cuts have been devised, which essentially
demand a pair of forward--backward pointing jets with an invariant mass of
$M_{\Pj\Pj}\gsim500\GeV$ and a large rapidity gap of $|\Delta y_{\Pj\Pj}|\gsim2.5$.
These cuts strongly suppress the VVV background
and enhance the EW VBS contribution to like-sign WW~scattering to about $90\%$,
but the QCD part in the other VBS channels still remains about $80\%$.

\begin{figure}
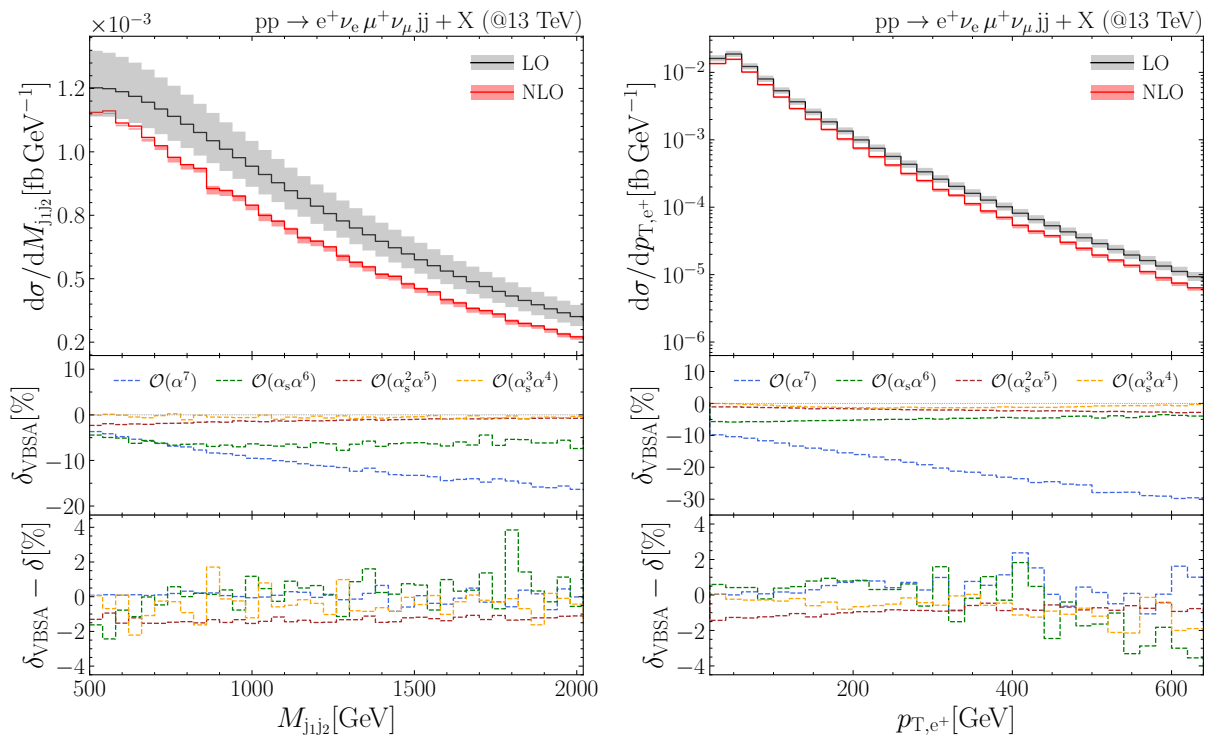

\vspace{1.5em}
\includegraphics[bb= 10 3 458 557,clip,page=7,width=0.49\textwidth]{nlos_vs_vbsa} \hfill
\includegraphics[bb= 10 3 458 557,clip,page=8,width=0.49\textwidth]{nlos_vs_vbsa}
\caption{Distributions in the di-jet invariant mass and the 
transverse momentum of the positron for like-sign WW~scattering at the LHC:
absolute predictions (top panels), relative NLO corrections in VBSA (middle panels),
and difference between relative full NLO corrections and VBSA (bottom panels).
[Taken from \protect \citere{Dittmaier:2023nac}.]}
\label{fig:VBS-Mjj-pTl}
\end{figure}

The NLO corrections to the cross sections comprise contributions of the orders
${\cal O}(\alpha^7)$, ${\cal O}(\alphas\alpha^6)$,
${\cal O}(\alphas^2\alpha^5)$, and ${\cal O}(\alphas^3\alpha^4)$.
While the QCD corrections to the EW and QCD channels are known already for about 20~years
for all relevant VBS channels
(see \citere{Huss:2025nlt} for references),
the NLO towers including EW corrections
(up to very few remaining, less prominent orders)
have been completed in recent years for
$\PW^\pm\PW^\pm$~\cite{Biedermann:2016yds,Dittmaier:2023nac},
$\PW\PZ$~\cite{Denner:2019tmn},
$\PZ\PZ$~\cite{Denner:2020zit}, and
$\PW^\pm\PW^\mp$~\cite{Denner:2022pwc} scattering.
The NLO corrections to all VBS channels share the feature that the purely EW corrections,
i.e.\ the ones of ${\cal O}(\alpha^7)$, are quite large, typically $\sim-16\%$
normalized to the EW LO contribution that actually features EW VBS.
In Ref.~\cite{Biedermann:2016yds} this large effect was traced back to the impact
of EW Sudakov corrections $\propto(\alpha/\sw^2)\ln^2(Q^2/\MW^2)$ plus subleading
single-logarithmic universal corrections, where $Q\sim400\GeV$ corresponds to the
mean invariant mass of the produced four-lepton system.
All other NLO orders are less prominent. 
The l.h.s.\ of \refta{tab:VBSvsVVV} illustrates the LO contributions
(after applying typical VBS cuts), the relative NLO corrections, and the residual
scale uncertainties for the integrated cross section of a typical channel of
like-sign WW~scattering;
\reffi{fig:VBS-Mjj-pTl} shows two corresponding distributions.%
\footnote{For like-sign WW~scattering, results on the matching of parton showers 
to the fixed-order NLO corrections can be found in 
\citeres{Melia:2010bm,Jager:2011ms,Chiesa:2019ulk,Jager:2024eet}.}
\begin{table}
\caption{Comparison of relative LO contributions $\Delta_{\EW/\QCD}$,
relative corrections $\delta^{\EW/\QCD}$, 
and residual scale dependences 
of integrated cross sections
for $\Pep\nu_\Pe\mu^+\nu_\mu\Pj\Pj$ (like-sign~WW)
productions, analyzed with VBS or VVV event selections.}
\vspace{.5em}
\centerline{\setlength{\tabcolsep}{10pt}
\def\arraystretch{1.2}
\begin{tabular}{l|l|l}
\hline
& VBS (based on \citeres{Biedermann:2016yds,Dittmaier:2023nac})
& VVV (based on \citere{Denner:2024ufg})
\\
\hline
selection cuts &
$M_{\Pj\Pj} > 500\GeV$, \quad $\Delta y_{\Pj\Pj} > 2$ &
$M_{\Pj\Pj} < 160\GeV$, \quad $\Delta y_{\Pj\Pj} < 1.5$
\\
& $M_{\ell\ell} > 20\GeV$, ...  & $40\GeV < M_{\ell\ell} < 400\GeV$, ...
\\
LO contributions & $\Delta_{\EW} \sim 85\%$  & $\Delta_{\EW} \sim 75\%$
\\
&                                 & \quad $\Delta_{\PW\PW\PW} \sim 53\%\,\Delta_{\EW}$
\\
&                                 & \quad $\Delta_{\PW\PH} \sim 41\%\,\Delta_{\EW}$
\\
& $\Delta_{\QCD} \sim 13\%$  & $\Delta_{\QCD} \sim 25\%$
\\
\hline
EW corrections $\delta^{\EW}_{q\bar q...}$ & {$\sim-16\%\,\Delta_{\EW}+\dots$} & {$\sim-7\%\,\Delta_{\EW}+\dots$}
\\
\phantom{EW corrections} $\delta^{\EW}_{q\gamma...}$ & $\sim \phantom{x}{+}3\%\,\Delta_{\EW}$ & $\sim +3\%\,\Delta_{\EW}$
\\
\hline
QCD corrections $\delta^{\QCD}$  & {$\sim -7\%\,\Delta_{\EW}$} & {$\sim +39\%\,\Delta_{\EW}$}
\\
& $\phantom{\sim} -2\%\,\Delta_{\QCD}+\dots$ & $\phantom{\sim} +87\%\,\Delta_{\QCD}+\dots$
\\
\hline
residual scale dependence
& $\LO\sim10\%$, \quad $\NLO\sim4\%$ &
$\LO\sim5\%$, \quad $\NLO\sim6\%$
\\
\hline
\end{tabular} }
\label{tab:VBSvsVVV}
\end{table}%
For VBS processes the residual scale dependence typically reduces from $\sim10\%$
to $\sim4\%$ in the transition from LO to NLO.

The enormous complexity of NLO calculation for $2\to6$ particle processes like VBS
calls for proper approximations, e.g.\ to render predictions in SM extensions less costly.
In this spirit, early calculations of QCD corrections to the EW contribution had been
performed in ``$t$-channel approximation'', which is justified by the strong suppression
of gluon exchange between the two quark lines owing to colour conservation
(see, e.g., \citere{Ballestrero:2018anz}).
In a similar spirit, Ref.~\cite{Dittmaier:2023nac} discusses a ``VBS approximation (VBSA)''
that neglects all VVV contributions, gluon-fusion contributions (if there are any),
and corrections to LO interference diagrams between $t$- and $u$-channels.
Moreover, the virtual corrections treat the
two produced vector-boson resonances in ``double-pole approximation''
(based on the leading terms in an expansion about the resonance poles), so that
the computational effort at NLO is enormously reduced w.r.t.\ the full off-shell calculation.
Figure~\ref{fig:VBS-Mjj-pTl} illustrates that the VBSA for like-sign WW~scattering is good
within $\lsim1.5\%$ in the experimentally relevant phase space.
For extremely high momentum transfer, of course, limitations of the VBSA are expected,
similar to the observations made for di- and tri-boson production.
\looseness-1

Finally, we comment on the old idea to evaluate VBS cross sections at high
energies in the ``effective vector-boson approximation'' 
(EVA)~\cite{Dawson:1984gx},
which extends the concept of partons inside hadrons to the case of weak vector bosons.
The vector-boson emission $q\to qV$ is approximated by its asymptotic behaviour
in the collinear limit, where it is logarithmically enhanced.
Older studies~\cite{Kuss:1995yv,Accomando:2006mc} have already indicated
that the approximative quality of the EVA is rather limited,
an observation that has been confirmed in the more recent extensive study of
\citere{Dittmaier:2023nac}, where several EVA variants were applied to
like-sign WW~scattering at LO.
Among the various technical subtleties in the EVA construction
(on-shell projection of the $WW\to WW$ kinematics, 
polarization states for incoming space-like W~bosons, 
Coulomb pole in the underlying $WW\to WW$ amplitude, etc.),
the event selection for VBS processes at hadron colliders
via tagging two jets turns out to be the show-stopper of the EVA,
because the transverse-momentum cuts on the tagging jets,
$p_{\rT,\Pj}>p_{\rT,\mathrm{cut}}$,
exactly exclude the collinear splitting region for $q\to qV$ which
is the basis for the EVA construction.
The results of \citere{Dittmaier:2023nac} demonstrate that the EVA 
is at best good at the qualitative level, and only if the forward/backward
regions of the jets are not excluded, but cannot serve as basis for
precise predictions.

\section{Electroweak tri-boson production}

EW tri-boson production is typically analyzed in fully leptonic final states ($6\ell$)
or in final states with four leptons and two jets ($4\ell2\Pj$). The former
provides the cleaner experimental signature, but the latter features larger
cross sections. Note also that the $4\ell2\Pj$ signature is the same as for VBS
processes, i.e.\ the same sets of Feynman diagrams are relevant for predictions,
but the experimental analyses focus on different kinematical regions.
Tri-boson analyses are carried out without applying VBS cuts, in order to
make the impact of VVV production diagrams more prominent.
VWW and VZZ production also receive contributions from VH production; in case of
VWW production the VH contribution cannot be suppressed if at least
one of the W~bosons resulting from the Higgs-boson resonance decays leptonically.

\begin{figure}
\includegraphics[width=0.52\textwidth,height=0.6\textwidth]{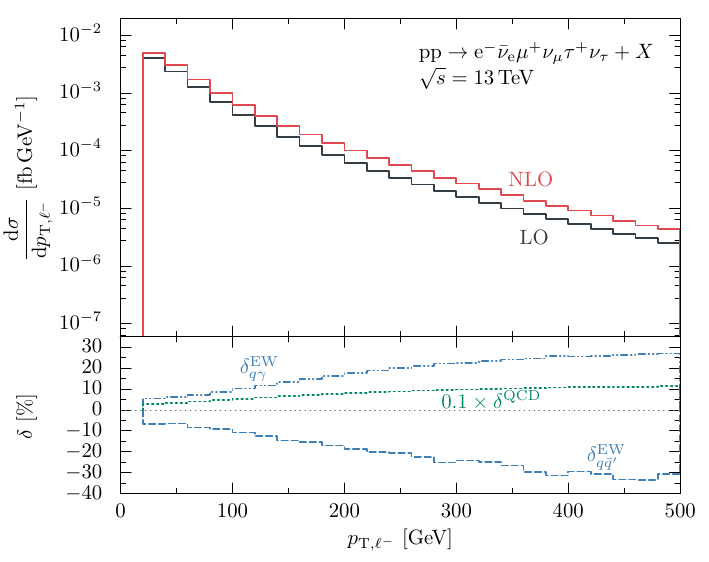}
\hfill
\includegraphics[width=0.46\textwidth]{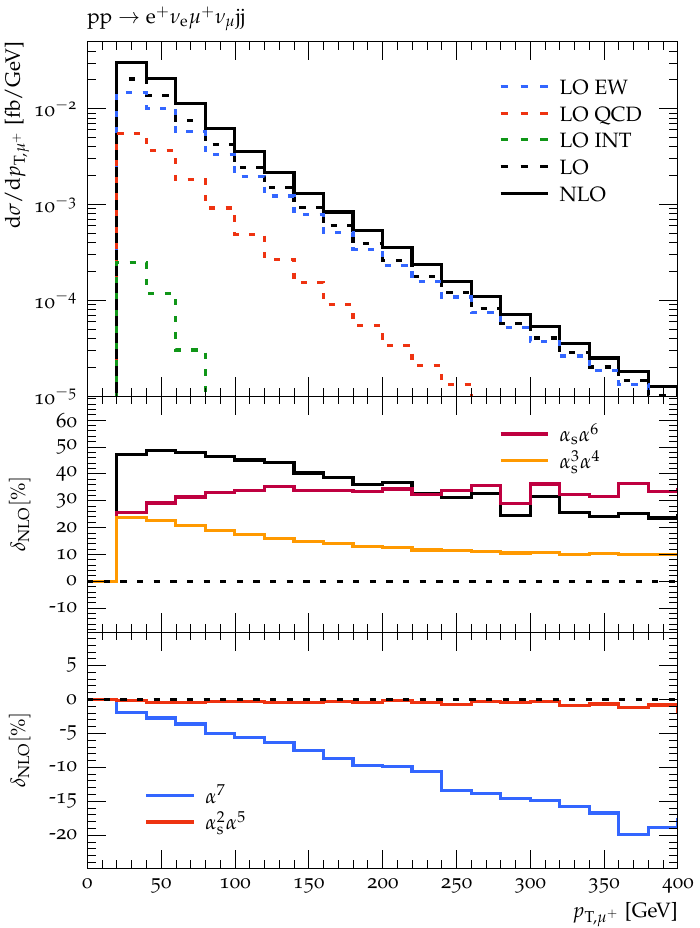}
\caption{Distributions in the transverse momentum of a charged lepton
in WWW~production with purely leptonically (left) and semi-leptonically (right)
decaying W~bsons.
On the left, the relative corrections of EW ($\delta^{\EW}_{q\bar q'}$) and
QCD ($\delta^{\QCD}$) origin as well as the $\gamma$-induced corrections
($\delta^{\EW}_{q\gamma}$) are shown in the lower panels;
on the right the lower panels show the relative corrections of the various orders 
$\alphas^m\alpha^n$.
[Taken from \protect \citeres{Dittmaier:2019twg,Denner:2024ufg}, respectively.]}
\label{fig:VVV-pTl}
\end{figure}

For fully leptonic final states, the LO VVV cross section scales like
${\cal O}(\alpha^6)$, so that genuine QCD and EW corrections of
${\cal O}(\alphas\alpha^6)$ and ${\cal O}(\alpha^7)$, respectively exist at NLO.
These corrections have been calculated in Refs.~\cite{Schonherr:2018jva,Dittmaier:2019twg}
for the full off-shell WWW processes.
While QCD corrections are about $\sim40\%$, the EW corrections receive contributions
of $\sim\mp(7{-}8)\%$ from $q\bar q$ and $q\gamma$ initial states to integrated cross
sections that tend to cancel against each other. The cancellation, however, is purely
accidental, strongly depends of phase-space cuts, and is less prominent in distributions,
where the EW corrections in the $q\bar q$ channels show the typical Sudakov enhancement
at high momentum transfer.
The l.h.s.\ of
\reffi{fig:VVV-pTl} illustrates the corrections for the distribution in the
transverse momentum of a charged decay lepton.
In addition to the NLO calculation for the full off-shell process,
a ``triple-pole approximation'' (TPA) for the NLO corrections based on the
contributions featuring a resonance enhancement by all three intermediate W~bosons
has been worked out and discussed in Ref.~\cite{Dittmaier:2019twg}.
For integrated cross sections and distributions in ranges with not too high
transverse momenta or invariant masses,
both the QCD and EW corrections are very well approximated by the TPA,
typically within $\lsim0.5\%$.
The quality of the TPA degrades in high-energy tails of distributions 
(like the missing energy carried away by neutrinos)
that are sensitive to ``collective recoil effects'' induced by contributions with less than
three W-boson resonances, but the quality of the TPA can be reliably assessed
from the impact on LO ``background'' contributions not featuring three resonant
W~bosons. 
Since VVV production in the fully leptonic channel does not involve strong-interaction
effects at LO, the residual scale dependence is small at LO and does not provide a
realistic estimate for theoretical uncertainties at LO; both at LO and NLO the
scale dependence is at the level of few percent.

For VVV production with a hadronically decaying vector boson,
the same tower of LO and NLO contributions as for VBS exists.
For some WWW and WWZ/WZZ production channels the NLO towers have been calculated
in Refs.~\cite{Denner:2024ufg} and \cite{Denner:2024ndl}, respectively
(at least partially in the latter case).
An exemplary result for WWW production is shown on the r.h.s.\ of
\reffi{fig:VVV-pTl} which illustrates the corrections for the distribution in 
the transverse momentum of a charged final-state lepton.
The figure reveals crucial differences to the corrections
found for the same final state in the kinematical VBS regime
(cf.~\reffi{fig:VBS-Mjj-pTl});
for integrated cross sections these differences are made more explicit in
\refta{tab:VBSvsVVV}.
Genuine EW corrections of ${\cal O}(\alpha^7)$ to the integrated cross section
are of the order of $\sim-7\%$, similar to the case of WWW production with leptonic
final states and less pronounced than in the VBS case.
The usual EW Sudakov enhancement for high momentum transfer drives these corrections to
$-(10{-}20)\%$ for transverse lepton momenta of $200{-}400\GeV$; for higher $p_{\rT}$
the cross section is suppressed by two orders of magnitude.
QCD corrections of ${\cal O}(\alphas\alpha^6)$ and ${\cal O}(\alphas^3\alpha^4)$ reach
several 10\%, rendering the inclusion of
NNLO QCD corrections desirable.
Finally, it should be mentioned that the cross section for WWW production shown 
on the r.h.s.\ of
\reffi{fig:VVV-pTl} contains an irreducible contribution of $\sim40\%$
originating from WH~production, so that the resulting values for the individual
NLO orders actually are a complicated mix of corrections to the genuine WWW,
WH, and VBS subprocesses.

\section{Conclusions}

The tremendous conceptual and technical progress in higher-order calculations
for many-particle processes
rendered NLO calculations possible for processes with 6--8 particles in the final state.
In particular, the full tower of ${\cal O}(\alphas^m\alpha^n)$ corrections is known
for the most important channels of EW VBS and EW tri-boson production processes.
The results, for instance, show generically large EW corrections of $\sim-16\%$
and $\sim-7\%$, respectively, for the EW contributions to the integrated cross
sections, and even larger corrections in high-energy tails of distributions,
rendering these corrections an important ingredients in predictions for
upcoming analyses of LHC data.

Leading-pole approximations, which employ expansions about W/Z~resonance poles,
as well as ``VBS approximations'', which neglect kinematically or colour-suppressed
interferences and VVV contributions in VBS cross sections, work sufficiently well
and, thus, provide a solid basis for studying
the polarizations of the W/Z~bosons
and for predictions within complicated
SM extensions, where full off-shell calculations might become too cumbersome.

\section*{Acknowledgments}

Thanks a lot to the organizers of Rencontres de Moriond for again
organizing this marvelous event in a stimulating atmosphere.
I am also grateful to Ansgar Denner, Gernot Knippen,
Philipp Maierh{\"o}fer,
Mathieu Pellen, Christopher Schwan and
Ramon Winterhalder
for the great collaboration on precision calculations for
EW multi-boson processes.


\section*{References}

\begin{thebibliography}{99}


\bibitem{Folgueras}
S.~Folgueras, these proceedings.

\bibitem{Han}
T.~Han, these proceedings.

\bibitem{Covarelli:2021gyz}
R.~Covarelli, \textit{et al.}
Int. J. Mod. Phys. A \textbf{36} (2021) no.16, 2130009
[arXiv:2102.10991 [hep-ph]].

\bibitem{Huss:2025nlt}
A.~Huss, \textit{et al.}
[arXiv:2504.06689 [hep-ph]].

\bibitem{Denner:2019vbn}
A.~Denner and S.~Dittmaier,
Phys. Rept. \textbf{864} (2020), 1-163
[arXiv:1912.06823 [hep-ph]].

\bibitem{Denner:2019tmn}
A.~Denner, \textit{et al.}
JHEP \textbf{06} (2019), 067
[arXiv:1904.00882 [hep-ph]].

\bibitem{Dittmaier:2019twg}
S.~Dittmaier, \textit{et al.}
JHEP \textbf{02} (2020), 003
[arXiv:1912.04117 [hep-ph]].

\bibitem{Denner:2022pwc}
A.~Denner, \textit{et al.}
JHEP \textbf{06} (2022), 098
[arXiv:2202.10844 [hep-ph]].

\bibitem{Dittmaier:2023nac}
S.~Dittmaier, \textit{et al.}
JHEP \textbf{11} (2023), 022
[arXiv:2308.16716 [hep-ph]].

\bibitem{Buccioni:2019sur}
F.~Buccioni, \textit{et al.}
Eur. Phys. J. C \textbf{79} (2019) no.10, 866
[arXiv:1907.13071 [hep-ph]].

\bibitem{Actis:2012qn}
S.~Actis,  \textit{et al.}
JHEP \textbf{04} (2013), 037
[arXiv:1211.6316 [hep-ph]];
%
Comput. Phys. Commun. \textbf{214} (2017), 140-173
[arXiv:1605.01090 [hep-ph]].

\bibitem{Denner:2016kdg}
A.~Denner, \textit{et al.}
Comput. Phys. Commun. \textbf{212} (2017), 220-238
[arXiv:1604.06792 [hep-ph]].

\bibitem{Denner:2026phn}
A.~Denner, \textit{et al.}
[arXiv:2602.19842 [hep-ph]].

\bibitem{Frederix:2018nkq}
R.~Frederix, \textit{et al.}
JHEP \textbf{07} (2018), 185
[erratum: JHEP \textbf{11} (2021), 085]
[arXiv:1804.10017 [hep-ph]].

\bibitem{Gleisberg:2008ta}
T.~Gleisberg, \textit{et al.}
JHEP \textbf{02} (2009), 007
[arXiv:0811.4622 [hep-ph]].

\bibitem{Biedermann:2016yds}
B.~Biedermann, \textit{et al.}
Phys. Rev. Lett. \textbf{118} (2017) no.26, 261801
[arXiv:1611.02951 [hep-ph]];
%
JHEP \textbf{10} (2017), 124
[arXiv:1708.00268 [hep-ph]].

\bibitem{Denner:2020zit}
A.~Denner, \textit{et al.}
JHEP \textbf{11} (2020), 110
[arXiv:2009.00411 [hep-ph]];
%
JHEP \textbf{10} (2021), 228
[arXiv:2107.10688 [hep-ph]].

\bibitem{Schonherr:2018jva}
M.~Sch{\"o}nherr,
JHEP \textbf{07} (2018), 076
[arXiv:1806.00307 [hep-ph]].

\bibitem{Denner:2024ndl}
A.~Denner, \textit{et al.}
JHEP \textbf{09} (2024), 187
[arXiv:2407.21558 [hep-ph]].

\bibitem{Melia:2010bm}
T.~Melia, \textit{et al.}
JHEP \textbf{12} (2010), 053
[arXiv:1007.5313 [hep-ph]].

\bibitem{Jager:2011ms}
B.~{J\"ager} and G.~Zanderighi,
JHEP \textbf{11} (2011), 055
[arXiv:1108.0864 [hep-ph]].

\bibitem{Chiesa:2019ulk}
M.~Chiesa, \textit{et al.}
Eur. Phys. J. C \textbf{79} (2019) no.9, 788
[arXiv:1906.01863 [hep-ph]].

\bibitem{Jager:2024eet}
B.~J{\"a}ger and S.~L.~P.~Chavez,
JHEP \textbf{01} (2025), 075
[arXiv:2408.12314 [hep-ph]].

\bibitem{Ballestrero:2018anz}
A.~Ballestrero, \textit{et al.}
Eur. Phys. J. C \textbf{78} (2018) no.8, 671
[arXiv:1803.07943 [hep-ph]].

\bibitem{Dawson:1984gx}
S.~Dawson,
Nucl. Phys. B \textbf{249} (1985), 42-60.

\bibitem{Kuss:1995yv}
I.~Kuss and H.~Spiesberger,
Phys. Rev. D \textbf{53} (1996), 6078-6093
[arXiv:hep-ph/9507204 [hep-ph]].

\bibitem{Accomando:2006mc}
E.~Accomando, \textit{et al.}
Phys. Rev. D \textbf{74} (2006), 073010
[arXiv:hep-ph/0608019 [hep-ph]].

\bibitem{Denner:2024ufg}
A.~Denner, \textit{et al.}
JHEP \textbf{08} (2024), 043
[arXiv:2406.11516 [hep-ph]].

\end{thebibliography}


\end{document}